\documentclass[letterpaper, 10 pt, conference]{ieeeconf}  

\IEEEoverridecommandlockouts                              
\usepackage{graphics} 
\usepackage{epsfig} 
\usepackage{mathptmx} 
\usepackage{times} 
\usepackage{amsmath} 
\usepackage{amssymb}  
\usepackage{float}

\usepackage{color}
\usepackage{eqparbox}

\usepackage[utf8]{inputenc} 
\usepackage[T1]{fontenc}    
\usepackage{booktabs}       
\usepackage{amsfonts}       
\usepackage{nicefrac}       
\usepackage{microtype}      
\usepackage{comment}

\usepackage[utf8]{inputenc} 
\usepackage[T1]{fontenc}    
\usepackage{hyperref}       
\usepackage{url}                
\usepackage{booktabs}       
\usepackage{amsfonts}       
\usepackage{nicefrac}       
\usepackage{microtype}      

\usepackage{comment}

\usepackage{amsmath,amssymb,amsfonts}
\usepackage{algorithmic}
\usepackage{algorithm}
\usepackage{graphicx}
\usepackage{textcomp}
\usepackage{xcolor}
\def\BibTeX{{\rm B\kern-.05em{\sc i\kern-.025em b}\kern-.08em
    T\kern-.1667em\lower.7ex\hbox{E}\kern-.125emX}}

\usepackage{color}

\begin{document}

\title{MIND: Model Independent Neural Decoder}

\author{ \parbox{1.5 in}{\centering Yihan Jiang\\
	ECE Department\\
         University of Washington\\
	Seattle, United States \\
	{\tt\small yij021@uw.edu}}
	 \parbox{1.5 in}{\centering Hyeji Kim\\
	Samsung AI Center Cambridge\\
         Cambridge, United Kingdom\\
	{\tt\small hkim1505@gmail.com}}
	 \parbox{1.5 in}{\centering Himanshu Asnani\\
	ECE Department\\
         University of Washington\\
         Seattle, United States \\
	{\tt\small asnani@uw.edu}}
	 \parbox{1.5 in}{\centering Sreeram Kannan\\
	ECE Department\\
         University of Washington\\
         Seattle, United States \\
	{\tt\small ksreeram@uw.edu}}
}





\maketitle

\begin{abstract}

Standard decoding approaches rely on model-based channel estimation methods to compensate for varying channel effects, which degrade in performance whenever there is a model mismatch. Recently proposed Deep learning based neural decoders address this problem by leveraging a model-free approach via gradient-based training. However, they require large amounts of data to retrain to achieve the desired adaptivity, which becomes intractable in practical systems.

In this paper, we propose a new decoder: Model Independent Neural Decoder (MIND), which builds on the top of neural decoders and equips them with a fast adaptation capability to varying channels. This feature is achieved via the methodology of Model-Agnostic Meta-Learning (MAML). Here the decoder: (a) learns a "good" parameter initialization in the meta-training stage where the model is exposed to a set of archetypal channels and (b) updates the parameter with respect to the observed channel in the meta-testing phase using minimal adaptation data and pilot bits. Building on top of existing state-of-the-art neural Convolutional and Turbo decoders, MIND outperforms the static 
benchmarks by a large margin and shows minimal performance gap when compared to the  neural (Convolutional or Turbo) decoders designed for that particular channel. In addition, MIND also shows strong learning capability for  channels not exposed during the meta training phase. 

\end{abstract}


 




\section{Introduction}\label{intro}

\subsection{Motivation}
Ever since the ground-breaking work in~\cite{shannon2001mathematical}, capacity-approaching codes for Additive White Gaussian Noise (AWGN) channel such as Turbo codes~\cite{berrou1993near}, Low Density Parity Check (LDPC) codes~\cite{mackay1996near} and Polar codes~\cite{arikan2008performance} have been proposed and extensively studied in the last few decades and have been used widely in Long Term Evolution (LTE) and 5G standards. Efficient decoding methods are known for the capacity-approaching codes, and they exhibit near-optimal performance on the Gaussian noise (AWGN) channel. However, the performance on non-AWGN channels is not uniformly optimal. Designing the corresponding decoders to deal with non-Gaussianity is hard, primarily owing to a two-fold deficit: (a) \textbf{model-deficit}, which implies the inability of accurately expressing the observed data by a clean mathematical model, and (b) \textbf{algorithm deficit}, which implies even under a clean abstraction, the optimal decoding algorithm is not known~\cite{jiang2018learn}. Thus while resorting to using the optimal codes designed under simplified models such as the AWGN channel, designing a decoder that can adapt to the non-AWGN channel effects faces challenges on both these fronts: there is a model mismatch and furthermore, most non-AWGN channels do not permit closed-form optimal decoders. 

Tremendous amount of effort has been invested to develop a suite of handcrafted algorithms to circumvent these deficits. These comprise of model-based methods in channel estimation, signal preprocessing, as well as robust decoding under unexpected channel effects~\cite{tse2005book}, so as to make the AWGN-designed capacity-approaching decoders operate with minimal degradation~\cite{richardson2008modern}. Few pilot bits known by both the transmitter and the receiver are used to estimate the channel effects to compensate for their varying nature, while handcrafted decoding algorithms have been applied to improve the decoder's robustness~\cite{richardson2008modern}. However they lack in two respects: (1) Channel estimation and channel-effect equalizing algorithms are model-based, hence when the underlying mathematical abstraction suffers from model-deficit, there is a suboptimal performance. (2) AWGN-designed decoders are not robust to unexpected and uncompensated noises. 

\subsection{Prior Art : Neural Decoding}
In the past decade, data-driven deep learning based methods have changed the landscape of several engineering fields such as computer vision and natural language processing, with revolutionary performance benchmarks~\cite{deng2009imagenet}~\cite{devlin2018bert}. Applying general purpose deep learning models to channel coding design has received intensive attention recently~\cite{o2016learning}~\cite{o2017introduction}. Designing such neural decoders naturally fits well with the data-driven supervised learning approaches, since both the received signals and the target messages can be simulated from the underlying encoder and channel models. In this way, both the model-deficit and algorithm-deficit are navigated by directly training a neural decoder on the sampled data. 

Designing neural decoder for several classes of codes such as LDPC codes, Polar codes and Turbo codes with versatile deep neural networks has seen a growing interest within the channel coding community. Imitating Belief Propagation (BP) algorithm via learnable neural networks shows promising performance for High-Density Parity-Check (HDPC) codes and LDPC codes~\cite{nachmani2016learning}~\cite{nachmani2018deep} and Polar codes~\cite{gruber2017deep}~\cite{cammerer2017scaling}. Near optimal performance of Convolutional Code and Turbo Code under AWGN channel is achieved via Recurrent Neural Networks (RNN) for arbitrary block lengths~\cite{kim2018communication}, which also shows robust and adaptive performance under non-AWGN setups. 
A further extension of RNN encoders (and decoders) reveal state-of-the-art performance for feedback channels~\cite{kim2018deepcode} and low latency schemes~\cite{jiang2018learn}. 
Thus while neural decoders show the promise of alleviating model and algorithm deficits,
compared to the traditional decoding methods which utilize limited amount of pilot bits to adapt, neural decoders require a huge amount of data ({\em information complexity}) and long computation time ({\em computational complexity}) to adapt to the new channel. This serious drawback renders them quite intractable and far from practical deployment. The relevant question we ask here is the following: \textbf{\textit{Can we design neural decoders that strengthen their adaptive property, so that only minimal re-training is necessary?}} In what follows, this question is investigated and answered in affirmative.

\subsection{Our Contribution}

We introduce meta learning to navigate the data-hungry nature of the neural decoder. Meta learning operates in two steps: (a) it firstly performs \textbf{meta training phase} by learning on a wide range of archetypal tasks, and then (b) during the \textbf{meta testing phase} enables learning new tasks faster, while consuming less adaptation data than learning from scratch~\cite{vanschoren2018survey}. Supervised meta learning has a natural connection to adaptive decoder design, as we can consider different channels as different tasks in our meta learning framework. 

RNN-based meta learning considers the whole meta learning approach as a large-scale RNN with tasks as inputs~\cite{santoro2016ml}. However, this requires complex modeling and thus shows degradation in performance with respect to scalability. Model Agnostic Meta Learning (MAML)~\cite{finn2017maml} is a gradient-based meta-learning algorithm that learns a sensitive initialization for fast adaptation. MAML trained model performs well on new tasks with limited gradient update steps and few-shot adaptation data. Compared to other meta learning methods, MAML has much less complexity. Moreover, theoretically MAML is shown to be able to approximate any meta learning algorithm~\cite{finn2018mamluniversal} and when faced with out-of-domain tasks, MAML shows fast capability to adapt, despite the fact that the out-of-domain tasks may not be close to the meta-trained tasks~\cite{finn2017maml}. 

In this work, we present a MAML-based neural decoder: Model Independent Neural Decoder (MIND), which admits fast adaptation with few shot adaptation data utilizing the gradient-based training. Compared to the adaptive neural decoders which require large amounts of gradient training steps and data to adapt to new channel settings, MIND can adapt to a new channel with small amount of pilot bits and few gradient descent steps. Compared to the traditional adaptive decoding method, MIND offers a model-free gradient-based meta-learning approach built on the top of neural decoders, resolving both the model and the algorithm deficit. Thus, MIND enhances the advantages of neural decoders with data and computational efficiency. 

The paper is organized as follows: Section \ref{mind} discusses the details of MAML which builds on the top of neural decoders to results in our proposed decoder: MIND. Section \ref{perf} analyzes the performance of MIND which shows very near-optimal performance with few shot adaptation data, under both trained and untrained channels. Section \ref{discussion} concludes with the scope and limitations of MIND and discussion on the future directions.

\section{Model Independent Neural Decoder}\label{mind}
We consider the two neural decoders for Convolutional Code and Turbo Code respectively~\cite{kim2018communication} to develop MIND (for details refer to the Appendix). Both these neural decoders have larger number of parameters compared to the traditional algorithms to deal with the issues of model deficit and algorithm deficit. However, training neural decoders till convergence requires large amounts of data. This leads to a slow adaptation with costly computations. 
In what follows, we propose the remedy through MAML, which are described below along with the choice of Loss function and the hyper-parameters:

\underline{\textit{Loss Function:}}

For neural decoders, the loss function is Binary Cross-Entropy (BCE) since decoder is a classification task. $f_\theta$ is the neural decoder with parameter $\theta$. Formally speaking, we are given a collection of $M$ training channels $\{T\} = \{T_i, i \in {1,...,M}\}$. For a specific channel $T_i$ with sampled received signal $x_i$ and target message $y_i$, the loss function associated with a particular channel $T_i$ can be represented as:

\begin{equation} \label{loss}
    L_{T_i} (f_\theta) = \sum_{x^{(j)}, y^{(j)} \sim T_i} BCE(f_\theta (x^{(j)}),  y^{(j)}).
\end{equation}

\underline{\textit{Meta Training Phase:}}

The meta training objective is to learn a sensitive initial weight for all the training channels. This operates as per the following two sub-steps:
\begin{itemize}
\item \textbf{Task Update:} For each channel $T_i$, MIND updates the model weights $\theta$ to $\theta_i'=\theta - \alpha \nabla_{\theta} L_{T_{i}}(f_\theta)$ with adaptation learning rate $\alpha$. This is called task update as the update for the parameter is done for each task, here channel. The updated weights $\theta'$ should learn themselves to be close to the optimal decoder for each channel $T_i$. 

\item \textbf{Meta Update:} Here, the goal is to do a meta update or to minimize the following loss for all training channels with respect to $\theta$:

\vspace{-1.0em}
\begin{equation} \label{loss_all}
   \min_{\theta}\sum_{i}  L_{T_i} (\theta_i')=  \min_{\theta} \sum_{i} L_{T_i} (\theta - \alpha \nabla_{\theta} L_{T_{i}}(f_\theta) )
\end{equation}
which via gradient descent with meta learning rate $\beta$, is equivalent to the following update:
\begin{equation}\label{loss_all_theta}
    \theta \leftarrow \theta - \beta \nabla_\theta \sum_{T_i \in \{T\} } L_{T_i}(f_{\theta_i'})
\end{equation}

Computing the above gradient is equivalent to computing the gradient of gradient of the BCE loss. Second order gradients as in Eq. (\ref{loss_all_theta}) are expensive. In this paper, we use First-Order MAML (FOMAML)~\cite{nichol2018reptile}, which treats higher order gradients as constant, thus ignoring the second-order terms. 
Note it is this step above which distinguishes such a training phase with the vanilla average learning, known as Multi-task Learning (MTL)~\cite{sener2016mt}, where instead of Eq. (\ref{loss_all_theta}) the following assignment via the average of gradients on all the channels is used:

\begin{equation}\label{loss_all_theta_mtl}
    \theta \leftarrow \theta - \beta \nabla_\theta \sum_{T_i \in \{T\} }  L_{T_i}(f_{\theta}).
\end{equation}
\end{itemize}

\underline{\textit{Meta Testing Phase:}}

During the meta testing phase, firstly pilot bits from the new channel $T_i$
are collected. Then the $\theta$ is updated via gradient descent $\theta' = \theta - \alpha \nabla_{\theta} L_{T_{i}}(f_\theta)$. 
MIND's meta training and testing phase is depicted in the appendix.

\textit{Note:} During the meta training phase, the data to compute task update $\nabla_\theta L_{T_i}(f_{\theta_i'})$ and the data for computing meta update  $\theta \leftarrow  \theta - \beta \nabla_\theta \sum_{T_i \in\{T\}} L_{T_i} (f_{\theta_i'})$ are different. Using the same data for both the task update and the meta update leads to meta-overfitting~\cite{finn2017maml}. It is due to this reason for training each $T_i$, we need to sample twice for meta training, while during the meta testing phase each step only requires to sample once.

\begin{figure}[!ht] 
\tiny
\centering
\vspace{-1.0em}
 \begin{tabular}{|c c c|} 
 \hline
 Parameters  & Convolutional Code & Turbo Code\\ [0.5ex] 
 \hline
 Neural Decoder & 2 layer bi-GRU  &2 layer bi-GRU  \\
 Number of Neural Units & 200  &200\\ 
 Batch Size $B$ & 100  &100\\ 
 Meta Batch Size $P$ & 10 &10\\
 Meta Learning Rate $\beta$ & 0.00001& 0.00001 \\
 Adaptation Learning Rate $\alpha$& 0.001 & 0.0001\\
 Number of Meta Update Steps & 50000 &50000\\   
 Block Length $L$& 100 &100\\
 Train SNR &  0 to 4dB & -1.5 to 2dB\\  
 Code Rate &  1/2 & 1/3\\  [1ex] 
 \hline
\end{tabular}
\vspace{-1.0em}
\caption{MIND Hyperparameters}\label{learn_hyper}
\vspace{-1.0em}
\end{figure}

\begin{figure}
\tiny
\centering
\begin{tabular}{ |c |c| c |}
\hline
 Testing Method & Adaptation Data & Task Update Steps $K$\\  
\hline
Fine-tune                   & 1000000  & 10000\\ 
MIND-1 Meta Testing & 100  & 1\\
MIND-10 Meta Testing & 1000   & 10\\
 \hline 
\end{tabular}
\caption{Adaptation cost between MIND and full adaptation}\label{finetune} 
\vspace{-3.0em}
\end{figure}

\underline{\textit{Hyperparameters:}}


The MIND trained Neural Decoders for Convolutional and Turbo Code are trained with the following hyper-parameters as shown in Figure \ref{learn_hyper}.  Batch size $B$ 
refers to the number of blocks sampled from one specific channel for training (also referred as mini-batch size), which is the same for both meta training and meta testing phase. Meta batch size $P$ refers to the number of random channels utilized for each meta training update step. Meta training is expensive, which uses 50000 training steps to conduct Meta Update. 
The adaptation rate $\alpha$ in the task update of meta training phase is larger than the meta learning rate $\beta$ of the meta update, which allows MIND to adapt faster. We use smaller adaptation learning rate $\alpha$ for neural Turbo decoder, due to its sensitive iterative decoding structure with shared model weights~\cite{kim2018communication}.

The data and computation cost for the meta testing phase is shown in Figure \ref{finetune}.
The task update step refers to the number of gradient steps $K$ required before testing on the new channel. Here we use the trained batch size $B=100$. Fine-tuning neural decoder without MIND to adapt to new channel requires $K=10000$ steps (each step need $B=100$ blocks) to converge. Compared to the fine-tuning, MIND only requires $K=1$ or $K=10$ gradient steps to conduct fast adaptation, with far less pilot data during the meta test phase.
In what follows for the evaluation of MIND's performance, MIND-$K$ refers to MIND with $K$ gradient update steps in the meta testing phase.


\section{MIND Performance}\label{perf}
In this section, we investigate the performance of MIND-$K$ for convolution code and turbo code against several benchmarks. 
\subsection{Channel Settings and Benchmarks}

The channels used in this paper are:
\begin{itemize}
	\item AWGN channel: $y = x+z$, $z \sim N(0, \sigma^2)$.
	\item Additive T-distribution Noise (ATN) channel:  $y = x+z$, where $z \sim T(\nu, \sigma^2)$. 
	\item Radar Channel: $y = x+z+w$. where $z \sim N(0, \sigma_1^2)$ is a background AWGN noise, and $w \sim N(0, \sigma_2^2)$, with probability $p$ is the radar noise with high variance and low probability. $\sigma_1 << \sigma_2$. 
\end{itemize}

\subsection{Benchmarks} 

For both the convolutional code and turbo code, we compare MIND-$K$ decoder against the following benchmarking decoders:
\begin{itemize}
\item \textbf{Canonical Optimal Decoders for AWGN Channel:} For convolutional code, Viterbi algorithm has optimal BER performance for AWGN channels~\cite{viterbi1967error}. For Turbo code, iterative Turbo decoder based on BCJR shows capacity-approaching performance. When decoding on AWGN channels, the above two decoders serve as useful benchmarks to be compared against.
\item \textbf{Adaptive Neural Decoders:} Under non-AWGN channels, generally there doesn't exist a close-form optimal decoding scheme. On the other hand, in these cases, neural decoders outperform most state-of-the-art heuristic decoders~\cite{kim2018communication}. Adaptive Neural Decoders are trained with nearly infinite data and computing resources on a particular channel and thus provide another useful benchmark especially for the non-AWGN channels.
\item \textbf{Multi-task Learning (MTL) based Decoders:} This is a benchmark for naive adaptation, termed as \textbf{MTL-$K$}, which updates weights via $K$-step gradient descent directly from MTL trained weight (Eq. \ref{loss_all_theta_mtl}),
with the same adaptation data batch size and learning rate as MIND-$K$. 
\end{itemize}

\subsection{MIND-$K$ for Convolutional Code}

\begin{figure}[!ht] 
\centering
\vspace{-1.0em}
\hspace*{-0.5cm}
\includegraphics[width=0.55\textwidth]{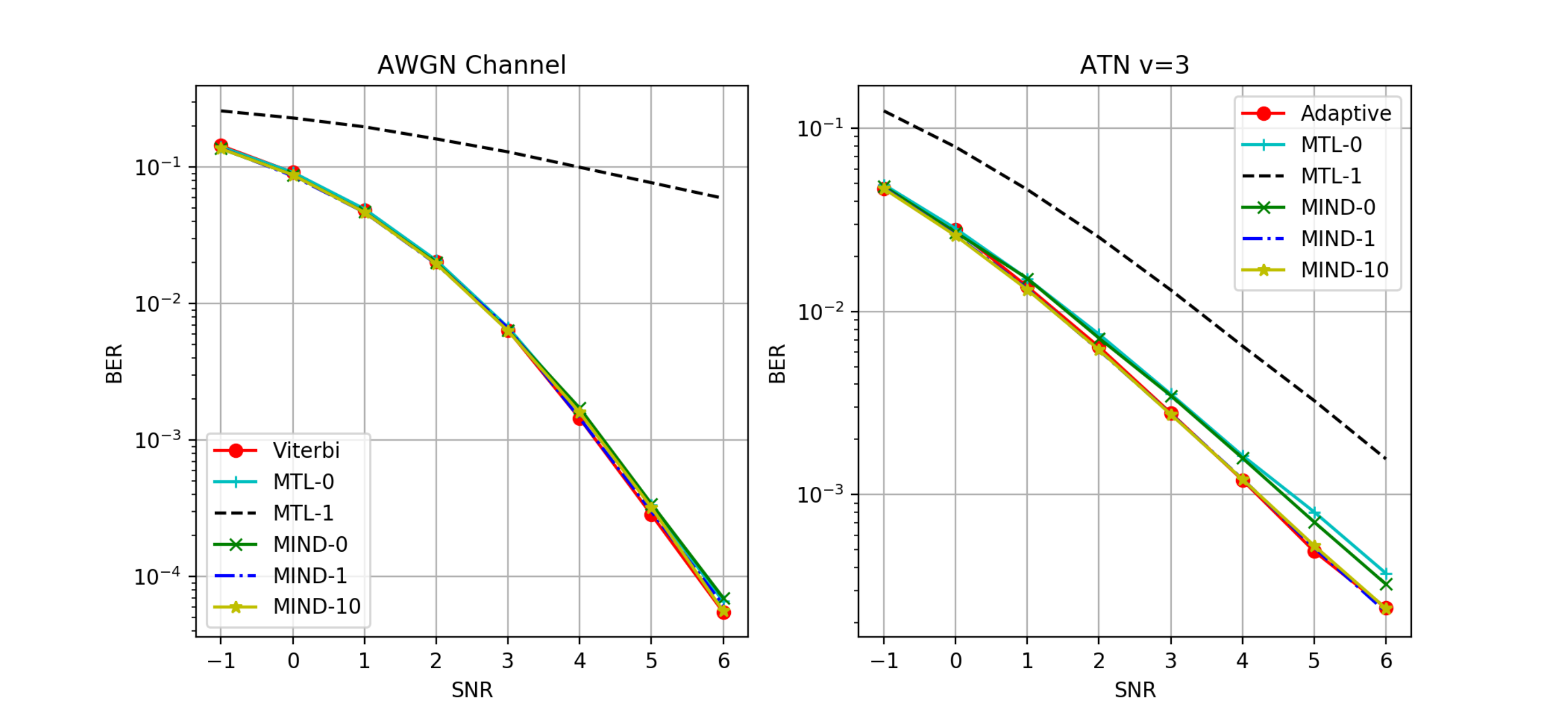}\ \ \ 
\hspace*{-0.5cm}
\includegraphics[width=0.55\textwidth]{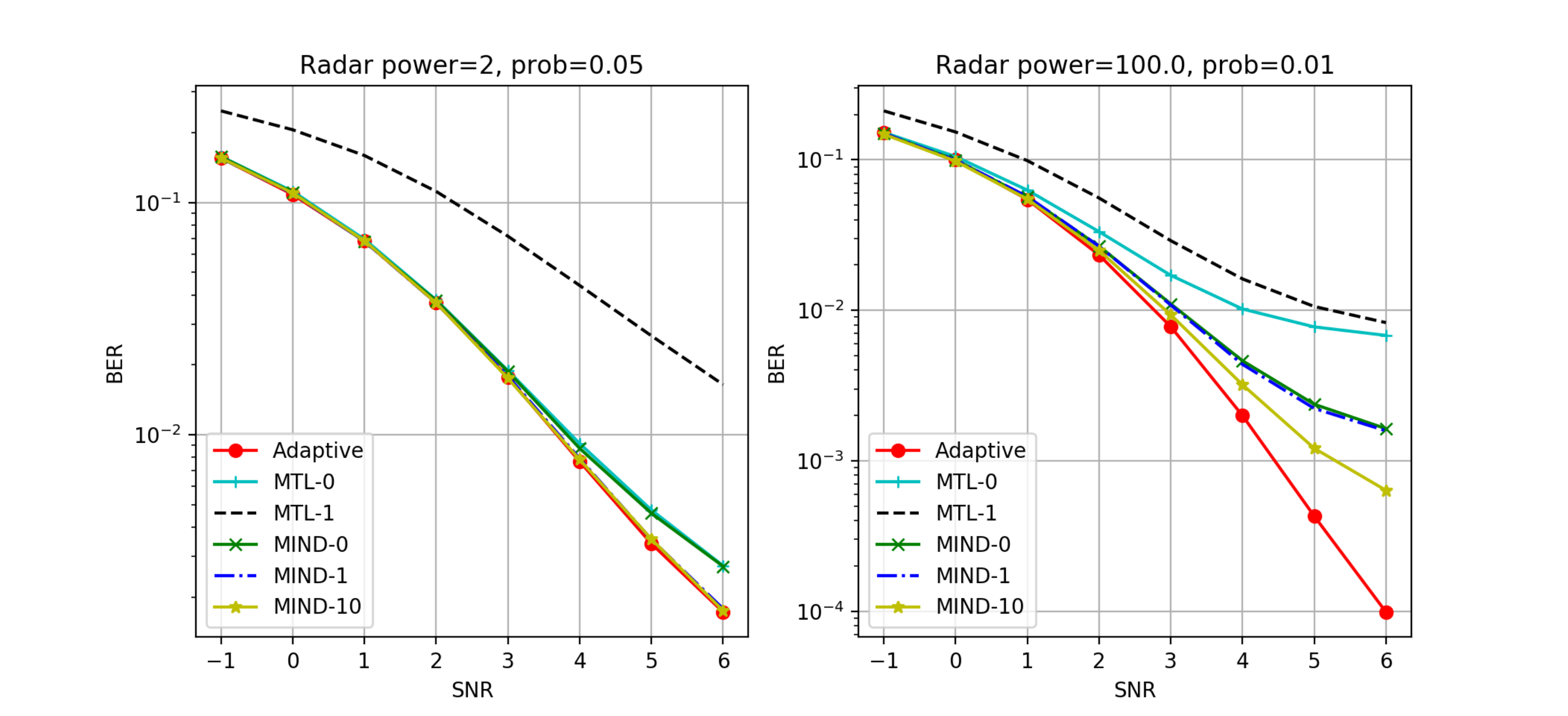}\ \ \ 
\vspace{-2.0em}
\caption{MIND for Convolutional Code: Trained AWGN (up left);Trained ATN ($\nu=3$) (up right);Trained Radar($\sigma_2=2.0, p=0.05$) (down left). and untrained Radar($\sigma_2=100.0, p=0.05$)(down right). 
\vspace{-1.0em}
}\label{maml_conv_test}

\end{figure}

We evaluate the fast adaptation ability under 4 different channels shown in Figure~\ref{maml_conv_test}: (1) AWGN channel, (2) Radar Channel ($\sigma_2=2.0$ and $p=0.05$), (3) ATN ($\nu=3.0$), and (4) untrained Radar ($\sigma_2=100.0, p=0.05$). The first three channels aim at testing the fast adaptation ability on meta-trained channels, where the fourth channel aims at testing learning ability on unexpected channel with dramatically different parameters. 

The MIND performance on Convolutional Code shows on trained channels:

\begin{itemize}
\item Among static methods without adaptation ability, MIND-0 and MTL-0 show similar performance. MIND without adaptation still performs well.
\item MIND-1 performs better than MTL-1, MIND-0, and MTL-0. MTL-1 shows a degradation indicating that a naive learning via average performance on all channels is not stable. 
\end{itemize}

To show the continued learning property on untrained channel, we also consider MIND-10 to compare. Here we observe:
\begin{itemize}
\item MIND-1 outperforms MTL-1, MTL-0, MIND-0. On untrained channel, MIND still shows improvement with solely gradient.
\item MIND-10 outperforms MTL-1.On untrained channel, apply more gradient steps can further improve performance.
\end{itemize}

\subsection{MIND-$K$ for Turbo Code}

As MTL-1 performs poorly, in this section we ignore MTL-1. On Turbo code, the channels tested shown below in Figure \ref{maml_turbo_test} are: (1) trained Radar channel ($\sigma_2=2.0, p=0.05$), and (2) untrained Radar channel ($\sigma_2=100.0, p=0.01$). 

\begin{figure}[!ht] 
\centering
\vspace{-1.0em}
\hspace*{-0.5cm}
\includegraphics[width=0.55\textwidth]{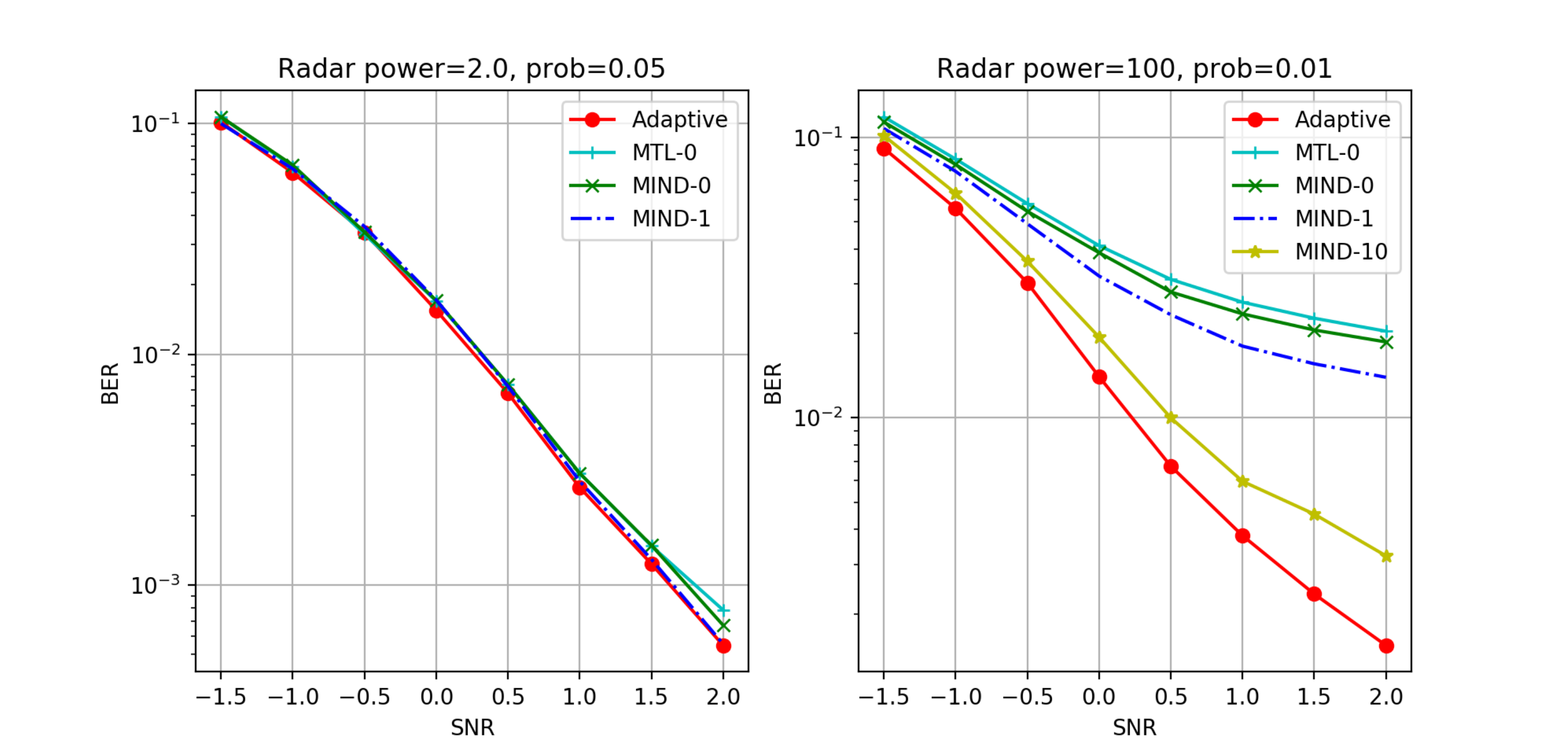}\ \ \ 
\vspace{-2.0em}
\caption{Neural Turbo Decoder with MIND. Trained Radar($\sigma_2=2.0, p=0.05$) (left), and untrained Radar($\sigma_2=100.0, p=0.01$)(right) }\label{maml_turbo_test}
\vspace{-1.0em}
\end{figure}

The performance on MIND with neural Turbo decoder shows the same trend as with Convolutional Code. The performance of MIND is consistent for both neural decoders as follows: 

\begin{itemize}
\item Without adaptation ability, MIND-0 shows robust performance, comparable to neural decoder trained on multiple channels.
\item With limited data and computation, MIND-1 outperforms static methods and shows performance close to optimal or adaptive algorithms.
\item On untrained channels, applying MIND with more gradient steps continually improves accuracy.
\end{itemize}

Comparing to deploying MTL-trained neural decoders, MIND shows comparable performance without adaptation ability, and can conduct fast adaptation with minimal re-training on both trained and untrained channels. For further detailed discussion as well as experiments on other channels, please refer to the appendix.

\section{Discussion}\label{discussion}






While we have designed MIND particularly for convolutional and Turbo codes, the methodology is not limited to these codes. In fact, the overall methodology is independent on the code structure or the neural network architecture, and thus can be adapted with equal felicity to other neural-based decoding problems. We note that MIND is not expected to be a universal decoder for \textbf{all} channels, rather that the learnt initialization is good for a class of channels which are related to the archetypal channels.
A precise characterization of this class is an interesting direction for future research. Furthermore, MIND still requires more samples than maybe available in a typical training channel. We expect neural method for joint channel estimation and data detection to perform better - this is left for future work.




Among future directions, it is worth considering to combine other neural decodes with MIND, such as neural LDPC~\cite{nachmani2016learning}~\cite{nachmani2018deep} and Polar~\cite{cammerer2017scaling} decoders. Beyond neural decoder design, MAML can also be applied to Channel Autoencoder~\cite{o2016learning} design, which deals with designing adaptive encoder and decoder. MAML is a growing area of interest in terms of its standalone research ~\cite{al2017continuous}~\cite{finn2018pmaml}~\cite{kim2018bmaml}~\cite{nichol2018reptile}~\cite{riemer2018cl}, with promising directions combining with online learning~\cite{finn2019onlinemeta}. These can usher new directions of remarkable improvements in decoder design.

\appendix


\subsection{Deep Learning Based Neural Decoders}\label{appendixdec}

In this section, we discuss the neural decoders for Convolutional Codes and Turbo Codes. We start with a small primer on Recurrent Neural Network (RNN) and Gated Recurrent Unit (GRU). 

\subsubsection{RNN and its variants}

\begin{figure}[!ht]
\centering
\includegraphics[width=0.38\textwidth]{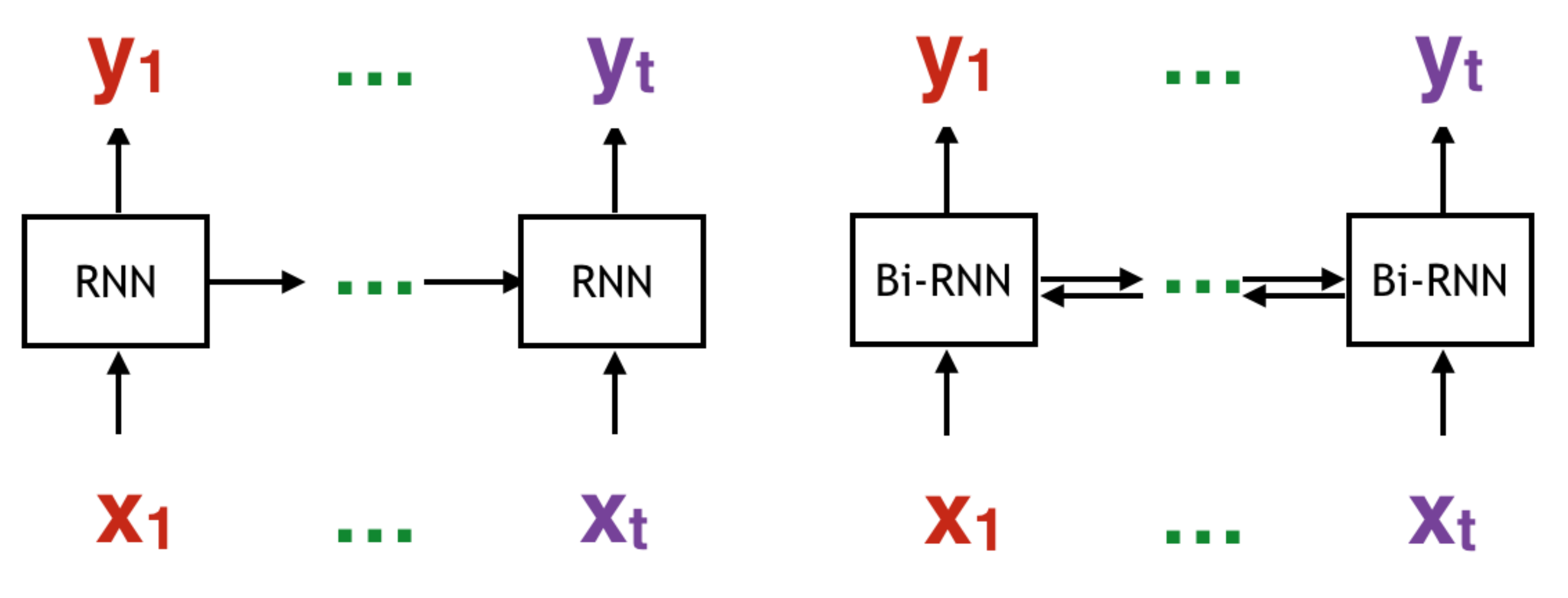}
\includegraphics[width=0.4\textwidth]{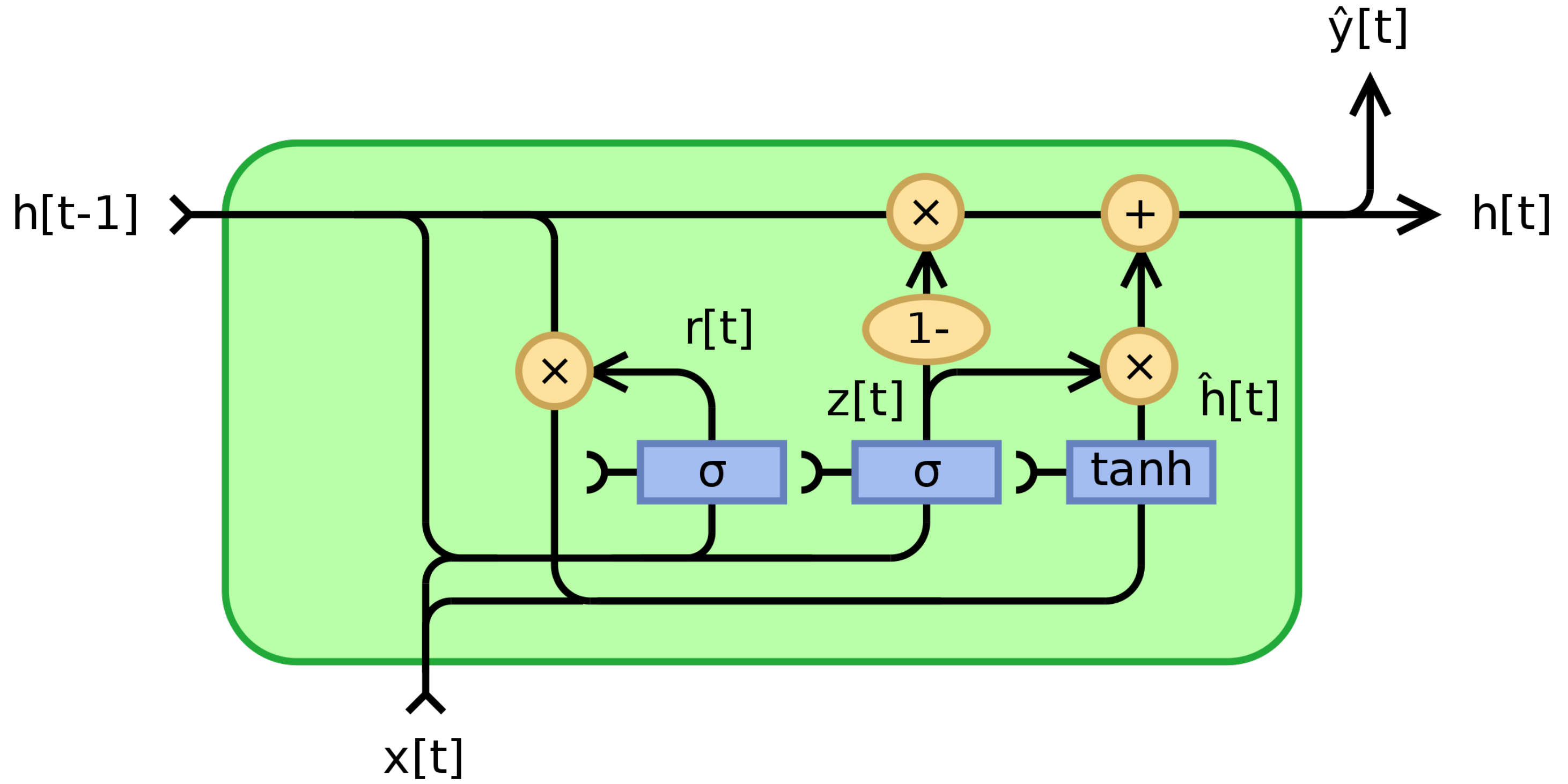}
\caption{RNN structure and bi-RNN (up), GRU (down)}
\label{rnns}
\end{figure}

Near-optimal neural decoders for both Convolutional Codes and Turbo Codes are based on Recurrent Neural Network (RNN)~\cite{kim2018communication}. RNN takes previous hidden state $h_{t-1}$ and $x_t$ as the input, and outputs the current output  $y_t$ and the current hidden state $h_{t}$ for the next time slot, defined as $(y_t, h_t) = f(x_t, h_{t-1})$. To use both the information from the past and the future, bidirectional RNN (bi-RNN) combines both forward and backward RNNs, defined as $(y_t, h^{f}_t, h^{b}_t) = f(x_t, h^{f}_{t-1}, h^{b}_{t+1})$. This is illustrated in Figure~\ref{rnns}~\cite{goodfellow2016deep}. 

Since vanilla RNN is hard to train due to exploding and diminishing gradients, Gated Recurrent Unit (GRU) is used as the primary network structure for neural decoders~\cite{chung2014empirical}. Bi-GRU uses the gating scheme as shown in Figure \ref{rnns} down, and is relieved of both the exploding and the diminishing gradients. In this paper, we use Bidirectional GRU (bi-GRU), GRU version of bi-RNN, as our primary neural structure.

\subsubsection{Neural Decoder for Convolutional Codes}

\begin{figure}[!ht]
\centering
\includegraphics[width=0.14\textwidth]{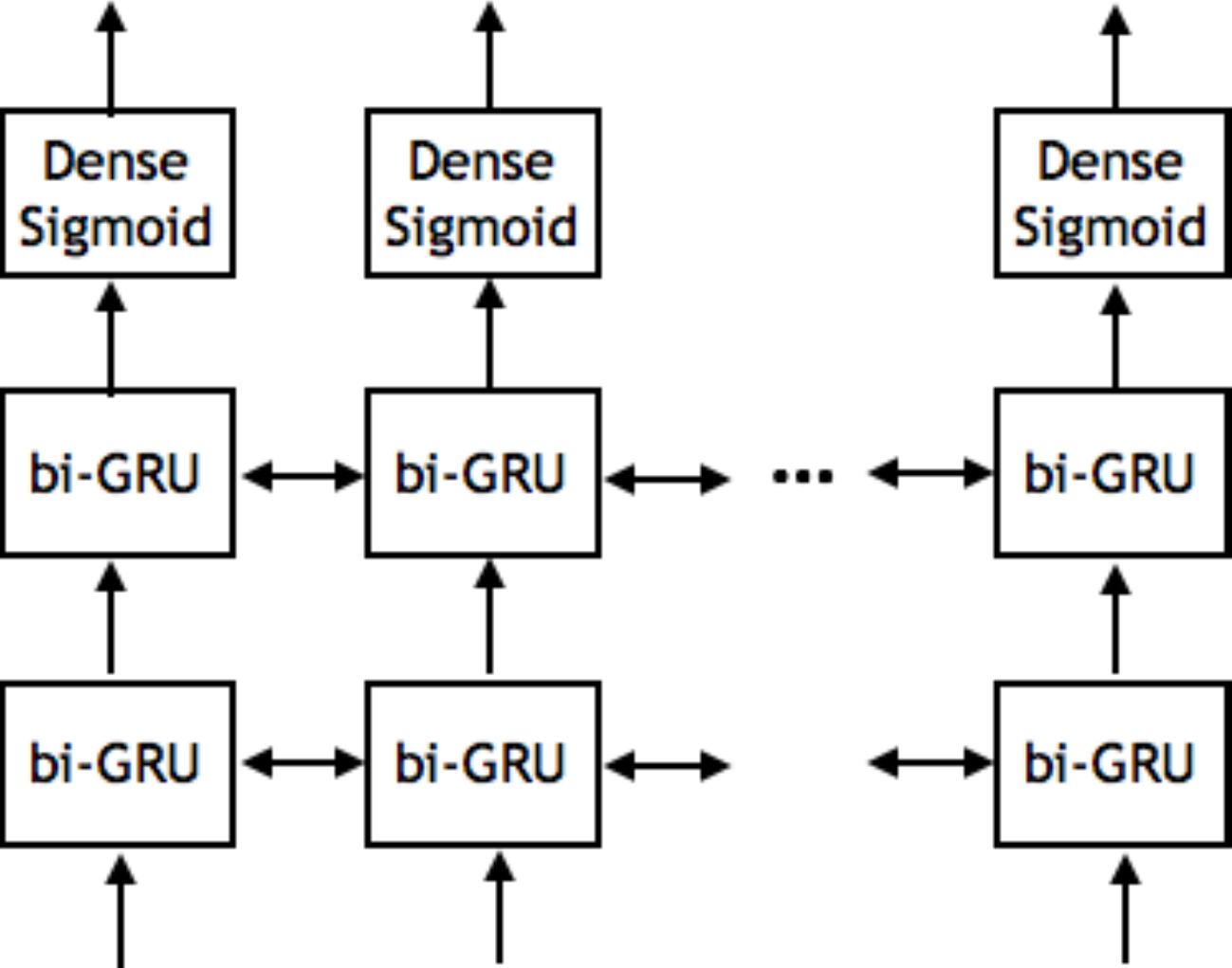}
\includegraphics[width=0.24\textwidth]{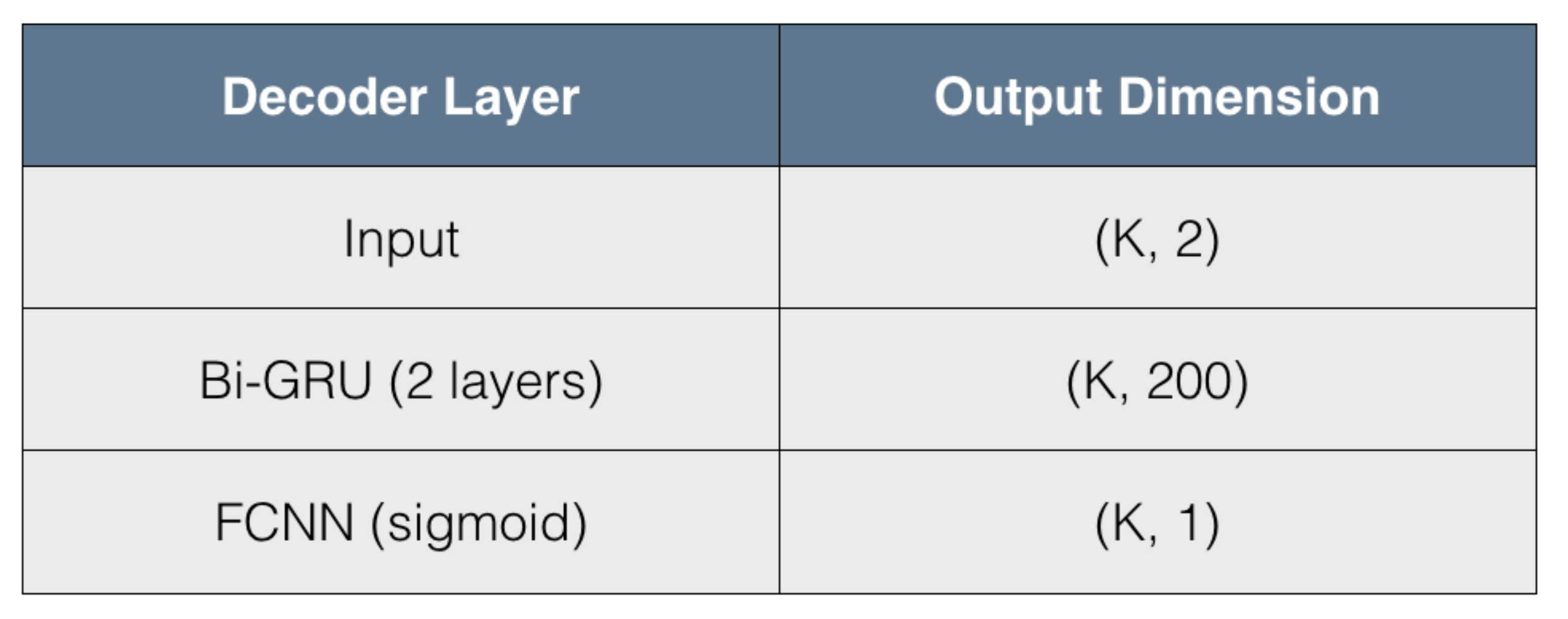} 
\caption{Neural decoder for Convolutional Codes (left), and network shape (right)}
\label{conv}
\end{figure}

Convolutional Code has theoretical optimal Viterbi and Bahl-Cocke-Jelinek-Raviv (BCJR) decoder under AWGN channel setting~\cite{viterbi1967error}~\cite{bahl1974optimal}. Inspired by the forward-backward structure of BCJR decoder, bi-GRU based neural decoder~\cite{kim2018communication} matches the optimal Block Error Rate (BLER) (BCJR) and  Bit Error Rate (BER) (Viterbi) performance under AWGN channel and outperforms existing heuristic algorithms under non-AWGN channels. More details about bi-GRU shown in appendix. The structure and hyper-parameter settings are shown in Figure \ref{conv} right two graphs. The convolutional encoder used in~\cite{kim2018communication} is a Recursive Systematic Convolutional (RSC) Code with generator sequence $f_1=[1 1 1]$ and $f_2 = [1 0 1]$, thus RSC encoder is represented by $[1, \frac{f_2}{f_1}]$.

\subsubsection{Neural Decoder for Turbo Codes}
\begin{figure}[!ht]
\centering
\includegraphics[width=0.45\textwidth]{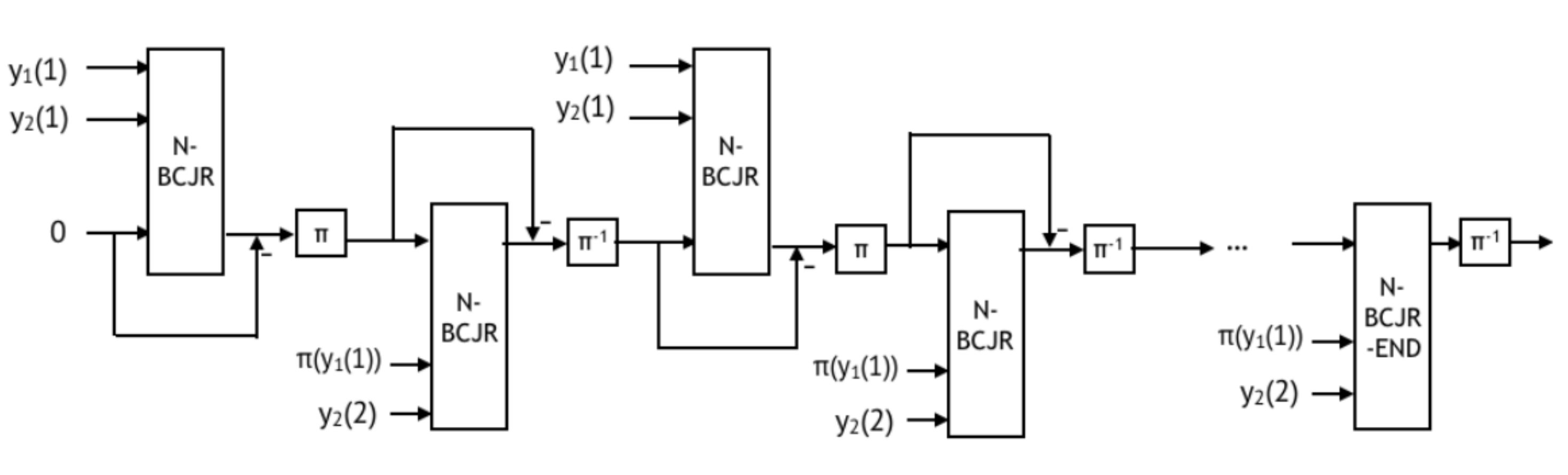}
\caption{Neural Turbo Decoder structure}
\label{ntd}
\end{figure}

Capacity-approaching Turbo code, as an extension of the convolutional code, can be near-optimally decoded by bi-GRU based neural decoders~\cite{kim2018communication}. The neural Turbo decoder structure is shown on Figure \ref{ntd}, where the N-BCJR blocks are pre-trained neural BCJR decoders with shared weights. 
Neural Turbo Decoder matches the performance of the state-of-the-art Turbo decoder under AWGN channel, while shows better performance under non-AWGN channels when compared to the widely-used heuristic algorithms~\cite{kim2018communication}. The Turbo code use the RSC encoder $[1, \frac{f_2}{f_1}]$, same as that for the Convolutional Code. The number of decoding iterations is 6. The neural N-BCJR decoder has the same design as shown in Figure \ref{conv} left third, except that the input shape changes to $(L,3)$ due to the inclusion of the likelihood bits. Further details regarding the implementation of RNN-based neural decoder can be found in~\cite{kim2018communication}.

\subsection{MIND Algorithm}\label{appendixmind}

\begin{algorithm}[H] 
  \caption{MIND : Meta-training Phase   \label{MAML1}} 
  \begin{algorithmic}[1] 
    \REQUIRE{\{T\} : training channel set}
    \REQUIRE{$\alpha$ adaption learning rate (task update), $\beta$ meta learning rate , $P$ meta batch size (meta update)}  
    \STATE{randomly initialize $\theta$}
    \STATE{\textbf{Task Update:}}
    \WHILE{not done}
      \STATE{Sample $\{T_P\}$ of $P$ channels from  $ \{T\} $}
      \FOR{$T_i \in \{T_P\}$ }
        \STATE{$\theta_i' = \theta $}
        \STATE{Sample $B$ data points $D = \{ x^{(j)}, y^{(j)} \}$ from $T_i$}
        \STATE{Compute $\nabla_\theta L_{T_i}(f_{\theta_i'})$ using $L_{T_i}$ in Equation (\ref{loss})}
        \STATE{Compute adapted parameters with gradient descent: $\theta_i' \leftarrow \theta_i' - \alpha \nabla_\theta L_{T_i}(f_{\theta_i'})$}
        \STATE{Sample another $B$ data points $D_i' = \{ x^{(j)}, y^{(j)} \}$ from $T_i$ for the meta-update}
      \ENDFOR
      \STATE{\textbf{Meta Update:}}
      \STATE{Update $  \theta \leftarrow \theta - \beta \nabla_\theta \sum_{T_i \in \{T\} } L_{T_i}(f_{\theta_i'})$}
    \ENDWHILE
  \end{algorithmic}
\end{algorithm}

\begin{algorithm}[H]\label{algo2}
  \caption{MIND : Meta Testing Phase  \label{MAML2}} 
  \begin{algorithmic}[1]  
    \REQUIRE{Channel $T_i$, Updates $K$}
    \REQUIRE{$\alpha$ adaption learning rate}
    \STATE{initialize with MIND meta trained $\theta'$ = $\theta$ (trained by Algorithm \ref{MAML1}}
    \FOR{$i \in 1,..., K$   }
        \STATE{Sample $B$ datapoints $D = \{ x^{(j)}, y^{(j)} \}$ from $T_i$}
        \STATE{Compute $\nabla_\theta L_{T_i}(f_{\theta_i'})$ using $L_{T_i}$ in Equation (\ref{loss})}
        \STATE{Update $\theta' \leftarrow \theta' - \alpha \nabla_\theta  L_{T_i} (f_{\theta'})$}
    \ENDFOR
    \STATE{Evaluate $L_{T_i}(f_{\theta_i'})$ using $L_{T_i}$ in Equation (\ref{loss})}
  \end{algorithmic}
\end{algorithm}

\subsection{MIND Performance}\label{appendixmindperf}
\subsubsection{MTL is hard and unstable to adapt}\label{CC}
\begin{figure}[H] 
\centering
\includegraphics[width=0.49\textwidth]{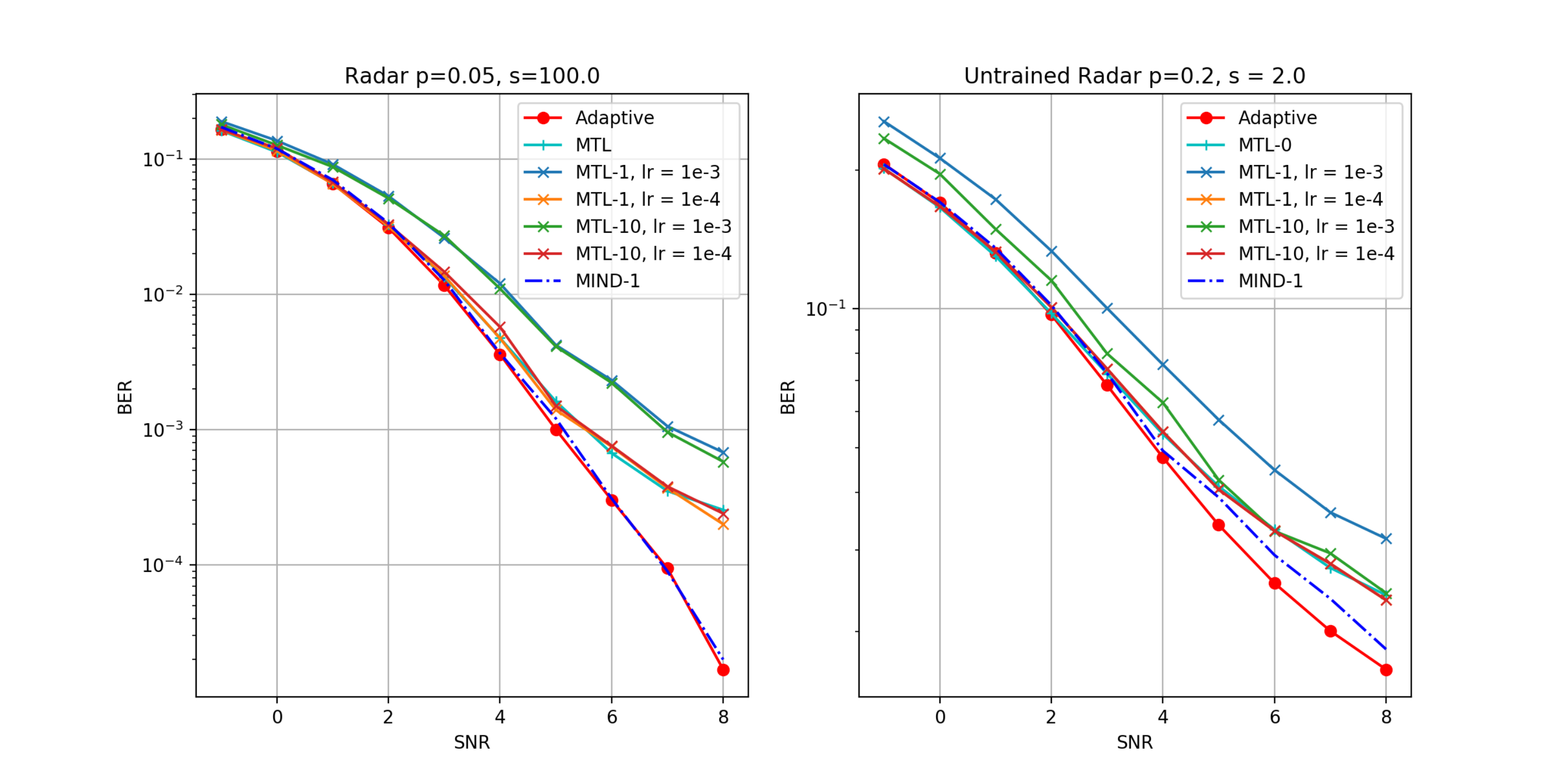}\ \ \ 
\caption{MTL with different $k$ and $\beta$: Trained Radar ($\sigma_2=2.0, p=0.05$)(left); Untrained Radar( $\sigma_2=2.0, p=0.2$)(right) }\label{mtl_naive}
\end{figure}

In section~\ref{perf}, we conducted MTL with the same adaption learning rate $\beta = 0.001$ as MIND. In this section we test MTL with different adaption learning rate $\beta=0.001, 0.0001$, with $K=1$ and $K=10$. Shown in Figure \ref{mtl_naive}, MIND-1 outperforms both MTL-1 and MTL-10 with different learning rates. Since MTL is not trained to conduct fast adaption, MTL-$K$ is very unstable. A naive application of the gradient descent on MTL leads to unstable and degrading performance. MIND learns to adapt with a large learning rate.

\subsubsection{MIND on Fading Channels}

\begin{figure}[H] 
\centering
\includegraphics[width=0.49\textwidth]{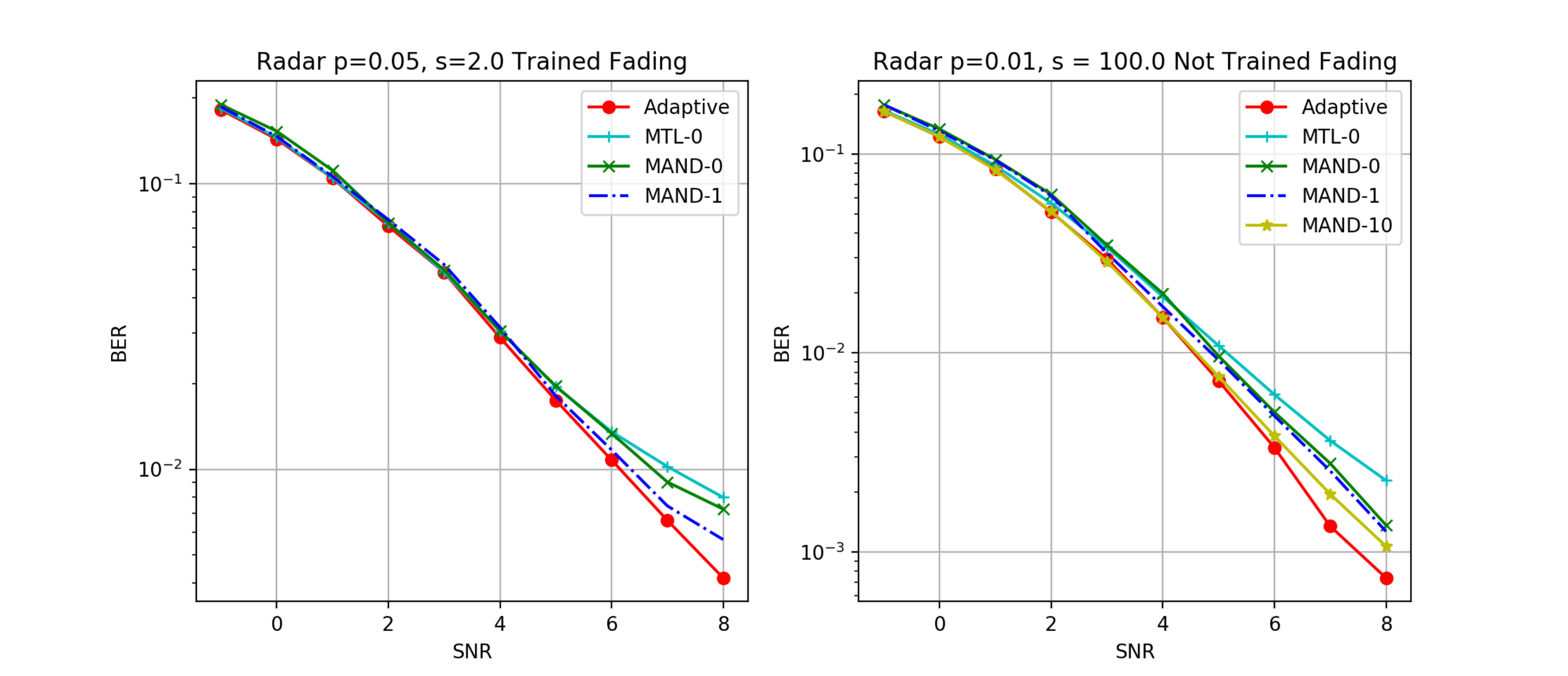}\ \ \ 
\caption{MIND for Convolutional Code on Fading channels: Trained Radar ($\sigma_2=2.0, p=0.05$)(left); Untrained Radar( $\sigma=100.0, p=0.01$)(right) }\label{maml_conv_fading}
\end{figure}

Fading channel can be represented as $y = hx + z$, where $h$ is the fading component, $z$ is the additive noise component as shown in the Section \ref{perf}. $h$ is taken to be normalized i.i.d Fast Rayleigh Fading, i.e. $h \sim \frac{\sqrt{X_1^2+X_2^2}}{\sqrt{\pi/2}}$, where $X_1$ and $X_2$ are independent standard Gaussian random variable. Further, normalizing with $\sqrt{\pi/2}$ gives $E(h) = 1$. We decode under coherent detection scheme when both $h$ and $y$ are fed to the decoder (inputs shape becomes $(L, 4)$ instead of $(L,2)$). When testing the adaption ability of additive noise channels on fading channels, the fading component is fixed. We test MIND under the same 2 different channels: (1) trained Radar channel ($\sigma_2=2.0, p=0.05$), and (2) untrained Radar Channel $\sigma_2=100$ and $p=0.01$, shown below with shown in Figure \ref{maml_conv_fading}. 

The result on fading channel shows same trend as in  non-fading scenario. On both channels, MIND-1 outperforms MIND-0 and MTL-0, and on untrained channel, MIND-10 outperforms MIND-1. 

\subsubsection{MIND on Diversified Training Channel Set}
In Section \ref{perf}, we meta-train the neural decoders with the following set of training channels:
\begin{itemize}
	\item AWGN channel.
	\item ATN with $v=5.0$ and $v=3.0$
	\item Radar with $p=0.05$, $\sigma_2=2.0, 3.5, 5.0$.
\end{itemize}

This is a somewhat less diversified training channel set, which contains closely distributed channels, which makes it possible to learn a neural decoder that works well on all the training channels. We want to test the performance on a more diversified training channel set. Towards this end, we use the following, which has channel parameters spanning a larger variation scale:
\begin{itemize}
	\item AWGN channel.
	\item ATN with $v=2.5$ and $v=3.0$
	\item Radar with $p=0.05, 0.2$, $\sigma_2=2.0, 10.0, 100.0$.
\end{itemize}

To test the learning ability of MIND under untrained channel, we use a testing channel set with both the trained channels mentioned above, and the following channels not in the training channel set, with more diversified channel parameters:
\begin{itemize}
	\item ATN with $v=10.0$ 
	\item Radar with $p=0.01, 0.1$, $\sigma_2=10.0, 100.0$.
\end{itemize}

\begin{figure}[H] 
\centering
\includegraphics[width=0.49\textwidth]{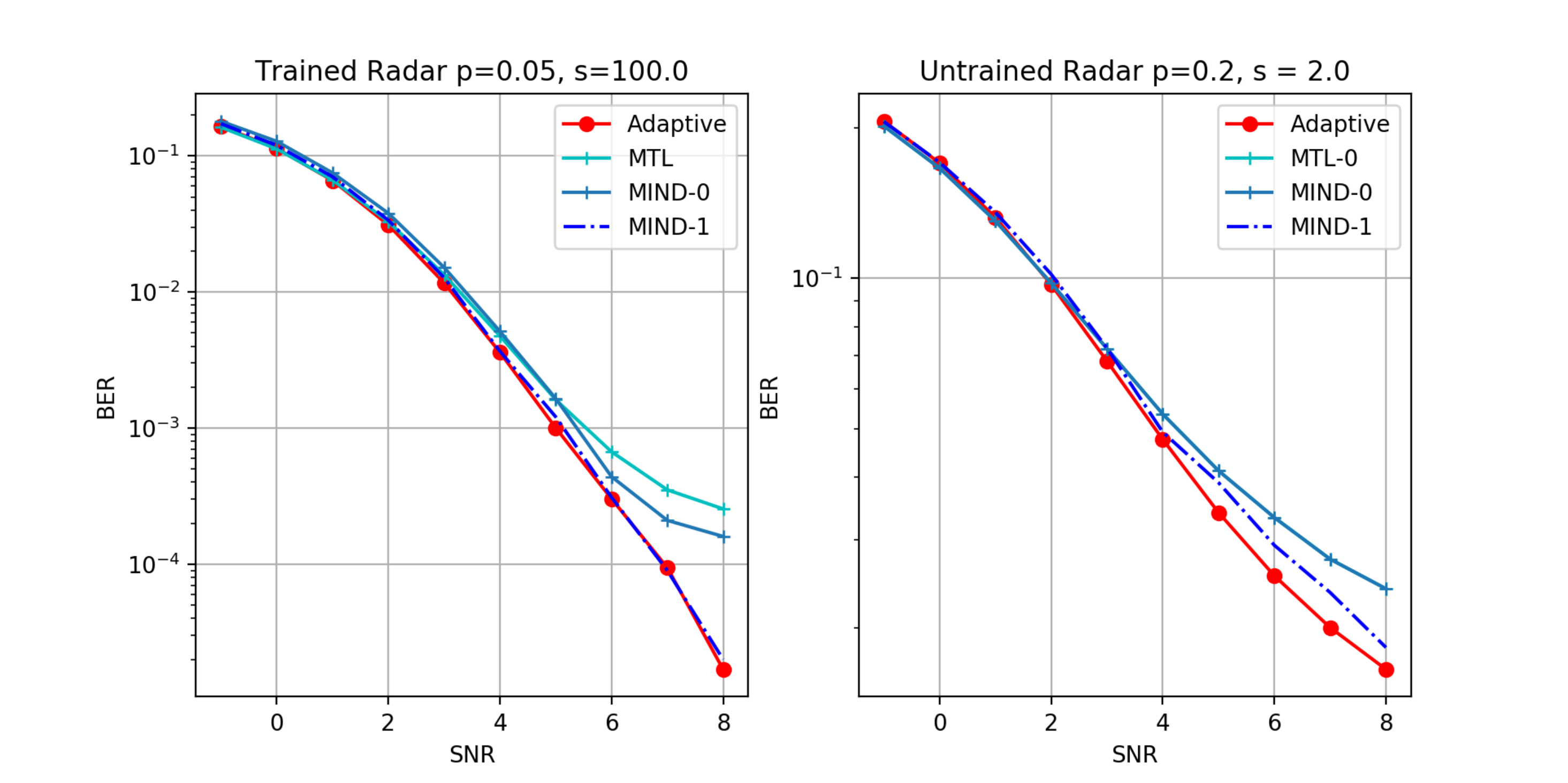}\ \ \ 
\caption{MIND trained with more diversified training set. Trained Radar ($\sigma_2=2.0, p=0.05$)(left); Untrained Radar( $\sigma_2=2.0, p=0.2$)(right) }\label{diverse}
\end{figure}

The performance on MIND trained with more diversified training channel set is shown in Figure \ref{diverse}. MIND still outperform MTL. MIND can handle training channel sets with a larger scale of diversity.

\subsection{Discussion on MIND Hyperparameters}\label{hyperparam}
We need to design MIND with proper hyper-parameters to control the trade-offs among data-efficiency, computations, and adapting stability. Three major hyper-parameters affect the performance of MIND: adaption batch size $B$, the test adaption steps $K$, and adaption learning rate $\alpha$. We empirically examine the effects of the above three hyper-parameters on neural Convolutional Code decoder as follows:

\subsubsection{Adaption batch size}
\begin{figure}[!ht] 
\centering
\includegraphics[width=0.5\textwidth]{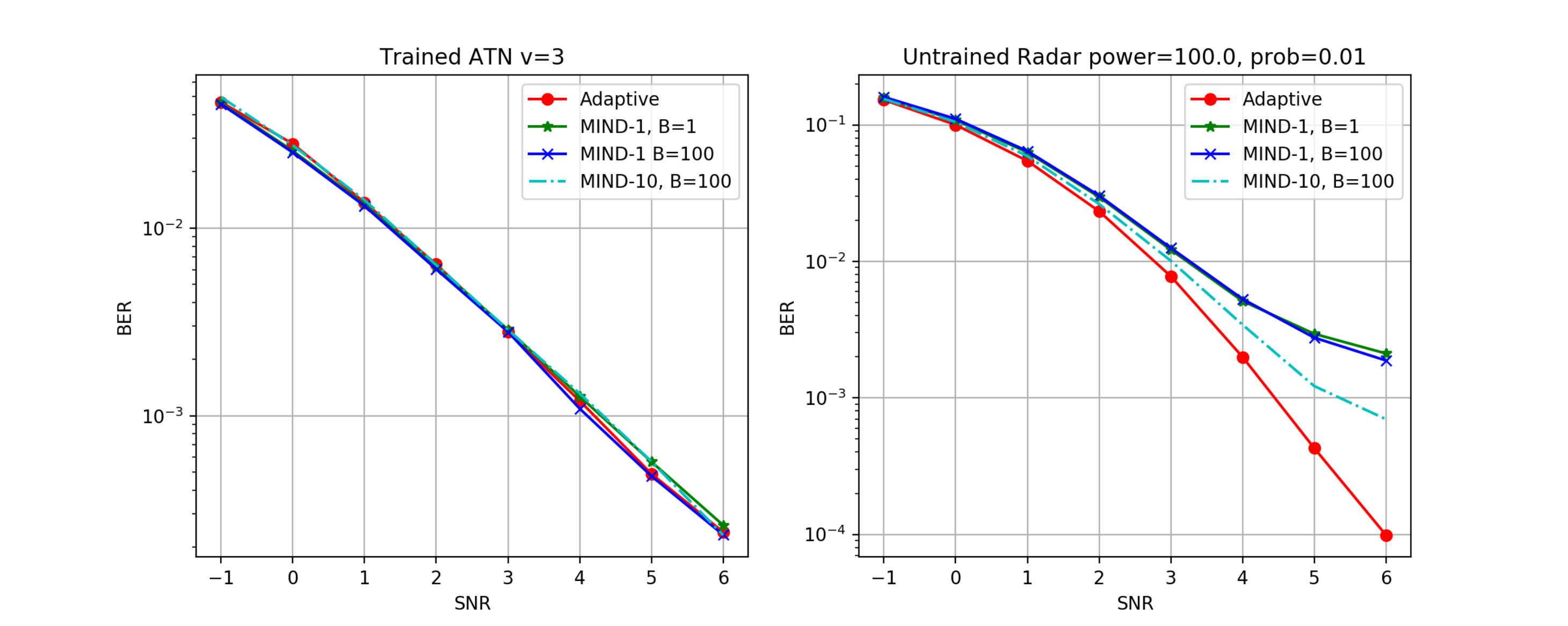}\ \ \ 
\caption{MIND adaption batch size $B$ and test adaption steps $K$ on trained ATN ($\nu=3.0$, middle), and untrained Radar ($\sigma_2=100.0, p=0.01$, right)}\label{maml_bs}
\end{figure}

The adaption batch size $B$ depends on the amount of available pilot data sampled from the new channel, which determines data-efficiency for MIND adaption. The performance between different adaption batch size is shown in Figure \ref{maml_bs}, which are trained on Radar channel ($\sigma_2=2.0, p=0.05$) and ATN channel($\nu=3.0$), and untrained Radar channel ($\sigma_2=100.0, p=0.01$).

On trained channels, different adaption batch sizes show similar performance close to optimal/adaptive methods. However on untrained ATN channel($\nu=3.0$), $B=100$ shows significantly improvement comparing to $B=1$. With small adaption batch size $B$, MIND trained model tends to only learn model which works well for all trained channels, without adaption ability. Only when the adaption batch size is large enough, MIND starts to utilize the data sampled from the new channel. Different adaption batch size $B$ reveals the trade-off between data-efficiency and adaption ability.

\subsubsection{Adaption steps}
The test adaption steps $K$ depends on the limitations of computation resources, which determine the computation efficiency for MIND. The effect of adaption steps $K$ is also shown in Figure \ref{maml_bs}. Note that on trained channel, adapting with $K=10$ steps and adapting with $K=1$ step show similar performance. However, on untrained channel, adapting with more steps improves the performance. The experiment shows that it is beneficial to conduct more adaption steps with MIND when testing on untrained channels.

\subsubsection{Adaption Learning rate}
\begin{figure}[!ht] 
\centering
\includegraphics[width=0.5\textwidth]{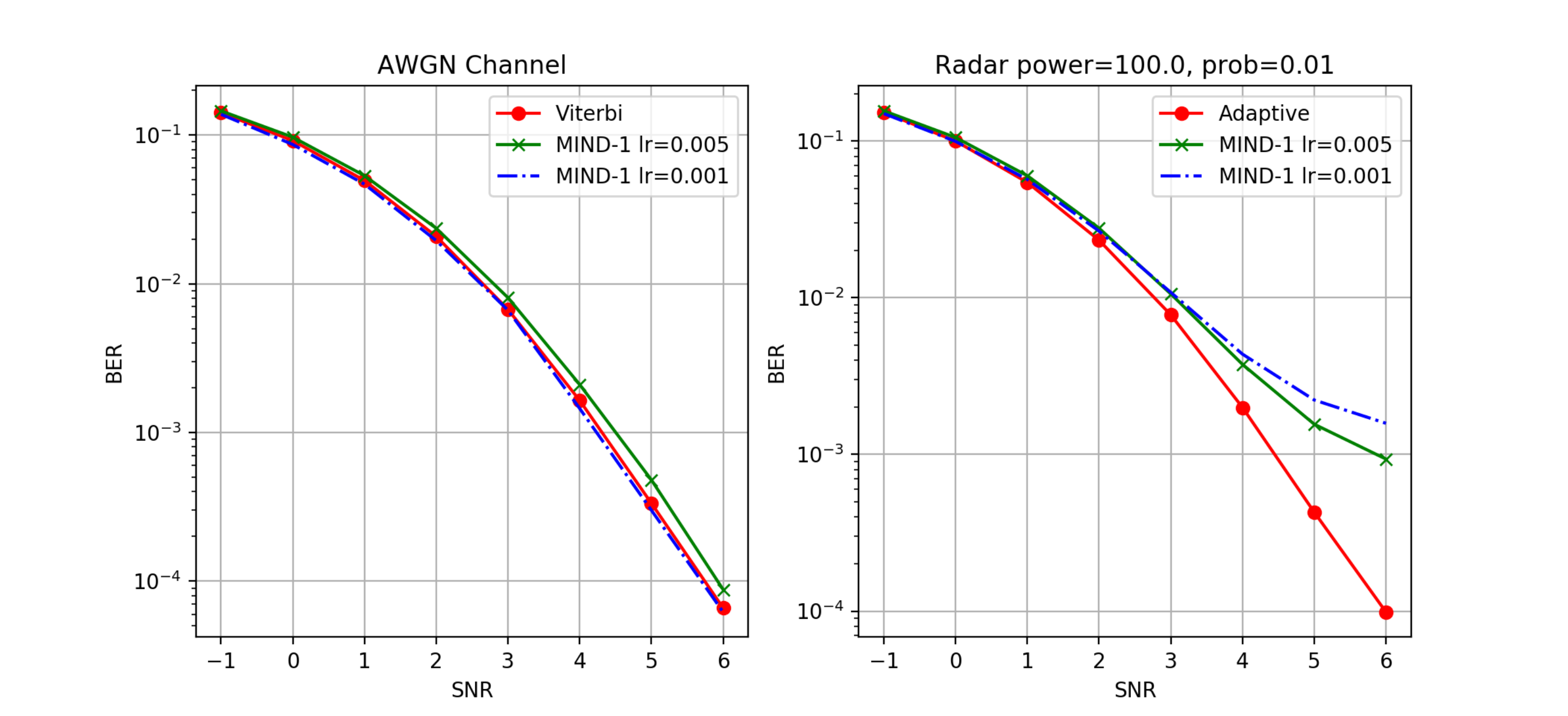}\ \ \ 
\caption{MIND adaption learning rate, AWGN channel (left), Radar channel (right)}\label{maml_lr}
\end{figure}

The adaption learning rate $\alpha$, controls the trade-off between stability and adapting speed. The performance of different adaption learning rate $\alpha$ is shown in Figure\ref{maml_lr}, on trained AWGN channel, and untrained Radar channel ($\sigma_2=100.0, p=0.01$). 

High adaption learning rate $\alpha=0.005$ shows worse performance on AWGN channel as shown in Figure\ref{maml_lr} left, while outperforms $\alpha=0.001$ on untrained Radar channel shown in Figure\ref{maml_lr} right. The experiment shows that adaption learning rate $\alpha$ controls the adapting aggressiveness of MIND. When trained with higher adaption learning rate $\alpha=0.005$, MIND learns to aggressively adapt with data from new channel, improves adapting ability on a new channel with sacrificing the performance on trained channels. On the other hand, with small adaption learning rate $\alpha=0.001$, MIND learns to conduct a somewhat conservative adaption.

Optimal adaption learning rate $\alpha$ depends on the use case. When the testing channel is similar to the training channel set, using smaller adaption learning rate is more favorable. When testing channel is very different comparing to training channel, a higher adaption learning rate is preferred.


\begin{thebibliography}{00}
\bibitem{tse2005book} Tse, David, and Pramod Viswanath. Fundamentals of wireless communication. Cambridge university press, 2005.
\bibitem{proakis1994comm} Proakis, John G. Communication systems engineering. Vol. 2. New Jersey: Prentice Hall, 1994.
\bibitem{sesia2011book} Sesia, Stefania, Matthew Baker, and Issam Toufik. LTE-the UMTS long term evolution: from theory to practice. John Wiley Sons, 2011.
\bibitem{vanschoren2018survey} Vanschoren, Joaquin. "Meta-learning: A survey." arXiv preprint arXiv:1810.03548 (2018).
\bibitem{riemer2018cl} Riemer M, Cases I, Ajemian R, Liu M, Rish I, Tu Y, Tesauro G. Learning to Learn without Forgetting By Maximizing Transfer and Minimizing Interference. arXiv preprint arXiv:1810.11910. 2018 Oct 29.
\bibitem{finn2017maml} Finn, Chelsea, Pieter Abbeel, and Sergey Levine. "Model-agnostic meta-learning for fast adaptation of deep networks."  International Conference on Machine Learning (ICML 2017)
\bibitem{finn2018mamluniversal} Finn, Chelsea, and Sergey Levine. "Meta-learning and universality: Deep representations and gradient descent can approximate any learning algorithm." International Conference on Learning Representations (ICLR 2018)
\bibitem{finn2018pmaml} Finn C, Xu K, Levine S. Probabilistic model-agnostic meta-learning. InAdvances in Neural Information Processing Systems 2018 (pp. 9537-9548).
\bibitem{kim2018bmaml} Yoon J, Kim T, Dia O, Kim S, Bengio Y, Ahn S. Bayesian model-agnostic meta-learning. InAdvances in Neural Information Processing Systems 2018 (pp. 7343-7353).
\bibitem{al2017continuous} Al-Shedivat M, Bansal T, Burda Y, Sutskever I, Mordatch I, Abbeel P. Continuous adaptation via meta-learning in nonstationary and competitive environments. International Conference on Learning Representations (ICLR 2018).
\bibitem{deng2009imagenet} Deng J, Dong W, Socher R, Li LJ, Li K, Fei-Fei L. Imagenet: A large-scale hierarchical image database. InComputer Vision and Pattern Recognition, 2009. CVPR 2009. IEEE Conference on 2009 Jun 20 (pp. 248-255). Ieee.
\bibitem{devlin2018bert} Devlin J, Chang MW, Lee K, Toutanova K. Bert: Pre-training of deep bidirectional transformers for language understanding. arXiv preprint arXiv:1810.04805. 2018 Oct 11.
\bibitem{santoro2016ml} Santoro A, Bartunov S, Botvinick M, Wierstra D, Lillicrap T. Meta-learning with memory-augmented neural networks. InInternational conference on machine learning 2016 Jun 11 (pp. 1842-1850).
\bibitem{shin2016tl} Hoo-Chang S, Roth HR, Gao M, Lu L, Xu Z, Nogues I, Yao J, Mollura D, Summers RM. Deep convolutional neural networks for computer-aided detection: CNN architectures, dataset characteristics and transfer learning. IEEE transactions on medical imaging. 2016 May;35(5):1285.
\bibitem{sener2016mt} Ruder S. An overview of multi-task learning in deep neural networks. arXiv preprint arXiv:1706.05098. 2017 Jun 15.
\bibitem{shannon2001mathematical} Shannon, Claude Elwood. "A mathematical theory of communication." Bell system technical journal 27.3 (1948): 379-423.
\bibitem{arikan2008performance} Arikan, Erdal. "A performance comparison of polar codes and Reed-Muller codes." IEEE Communications Letters 12.6 (2008).
\bibitem{berrou1993near} Berrou, Claude, Alain Glavieux, and Punya Thitimajshima. "Near Shannon limit error-correcting coding and decoding: Turbo-codes. 1." Communications, 1993. ICC'93 Geneva. Technical Program, Conference Record, IEEE International Conference on. Vol. 2. IEEE, 1993.
\bibitem{mackay1996near} MacKay, David JC, and Radford M. Neal. "Near Shannon limit performance of low density parity check codes." Electronics letters 32.18 (1996): 1645-1646.
\bibitem{richardson2008modern} Richardson, Tom, and Ruediger Urbanke. Modern coding theory. Cambridge university press, 2008.
\bibitem{farsad2018icassp} N. Farsad, M. Rao, and A. Goldsmith,  Deep learning for joint source-channel coding of text, IEEE International Conference on Acoustics, Speech and Signal Processing (ICASSP), 2018
\bibitem{li2013ofdma} Li, Junyi, Xinzhou Wu, and Rajiv Laroia. OFDMA mobile broadband communications: A systems approach. Cambridge University Press, 2013.
\bibitem{safavi2015impact} Safavi-Naeini, Hossein-Ali, et al. "Impact and mitigation of narrow-band radar interference in down-link LTE." Communications (ICC), 2015 IEEE International Conference on. IEEE, 2015.
\bibitem{goodfellow2016deep} Goodfellow I, Bengio Y, Courville A, Bengio Y. Deep learning. Cambridge: MIT press; 2016 Nov 18.
\bibitem{han2015deep} Han, Song, Huizi Mao, and William J. Dally. "Deep compression: Compressing deep neural networks with pruning, trained quantization and huffman coding." International Conference on Learning Representations (ICLR 2016).
\bibitem{hinton2006reducing} Hinton, Geoffrey E., and Ruslan R. Salakhutdinov. "Reducing the dimensionality of data with neural networks." science 313.5786 (2006): 504-507.
\bibitem{o2016learning} O'Shea, Timothy J., Kiran Karra, and T. Charles Clancy. "Learning to communicate: Channel auto-encoders, domain specific regularizers, and attention." Signal Processing and Information Technology (ISSPIT), 2016 IEEE International Symposium on. IEEE, 2016.
\bibitem{o2017introduction} O'Shea, Timothy, and Jakob Hoydis. "An introduction to deep learning for the physical layer." IEEE Transactions on Cognitive Communications and Networking 3.4 (2017): 563-575.
\bibitem{nachmani2016learning} Nachmani, Eliya, Yair Be'ery, and David Burshtein. "Learning to decode linear codes using deep learning." Communication, Control, and Computing (Allerton), 2016 54th Annual Allerton Conference on. IEEE, 2016.
\bibitem{nachmani2018deep} Nachmani E, Marciano E, Lugosch L, Gross WJ, Burshtein D, Be?ery Y. Deep learning methods for improved decoding of linear codes. IEEE Journal of Selected Topics in Signal Processing. 2018 Feb;12(1):119-31.
\bibitem{gruber2017deep} Gruber, Tobias, et al. "On deep learning-based channel decoding." Information Sciences and Systems (CISS), 2017 51st Annual Conference on. IEEE, 2017.
\bibitem{cammerer2017scaling} Cammerer, Sebastian, et al. "Scaling deep learning-based decoding of polar codes via partitioning." GLOBECOM 2017-2017 IEEE Global Communications Conference. IEEE, 2017.
\bibitem{kim2018communication} Kim, Hyeji and Jiang, Yihan and Rana, Ranvir and Kannan, Sreeram and Oh, Sewoong and Viswanath, Pramod. ''Communication Algorithms via Deep Learning'' Sixth International Conference on Learning Representations (ICLR 2018).
\bibitem{jiang2018learn} Jiang Y, Kim H, Asnani H, Kannan S, Oh S, Viswanath P. LEARN Codes: Inventing Low-latency Codes via Recurrent Neural Networks. arXiv preprint arXiv:1811.12707. 2018 Nov 30.
\bibitem{aoudia2018end} Aoudia, Faycal Ait, and Jakob Hoydis. "End-to-End Learning of Communications Systems Without a Channel Model." arXiv preprint arXiv:1804.02276 (2018).
\bibitem{felix2018ofdm} Felix A, Cammerer S, Dorner S, Hoydis J, Ten Brink S. Ofdm-autoencoder for end-to-end learning of communications systems. In2018 IEEE 19th International Workshop on Signal Processing Advances in Wireless Communications (SPAWC) 2018 Jun 25 (pp. 1-5). IEEE.
\bibitem{kim2018deepcode} Kim H, Jiang Y, Kannan S, Oh S, Viswanath P. Deepcode: Feedback codes via deep learning. InAdvances in Neural Information Processing Systems 2018 (pp. 9458-9468).
\bibitem{viterbi1967error} Viterbi, Andrew. "Error bounds for convolutional codes and an asymptotically optimum decoding algorithm." IEEE transactions on Information Theory 13.2 (1967): 260-269.
\bibitem{bahl1974optimal} Bahl L, Cocke J, Jelinek F, Raviv J. Optimal decoding of linear codes for minimizing symbol error rate (corresp.). IEEE Transactions on information theory. 1974 Mar;20(2):284-7.
\bibitem{chung2014empirical} Chung J, Gulcehre C, Cho K, Bengio Y. Empirical evaluation of gated recurrent neural networks on sequence modeling. arXiv preprint arXiv:1412.3555. 2014 Dec 11.
\bibitem{nichol2018reptile} Nichol, Alex, and John Schulman. "Reptile: a Scalable Metalearning Algorithm." arXiv preprint arXiv:1803.02999 (2018).
\bibitem{finn2019onlinemeta} Chelsea Finn, Aravind Rajeswaran, Sham Kakade, and Sergey Levine. "Online Meta-Learning." arXiv preprint arXiv:1902.08438 (2019)

\end{thebibliography}
\end{document}